\documentclass[bibnumber]{aa}

\usepackage{amsmath,amssymb}
\usepackage{graphicx}
\usepackage{bm}
\usepackage{booktabs}
\usepackage{hyperref}
\usepackage{xcolor}
\usepackage{multirow}
\makeatletter
\g@addto@macro\maketitle{\nolinenumbers}
\makeatother

\begin{document}
\nolinenumbers

\title{Non-Monotonic Rotation Imprint on Time-Integrated Neutrino Spectral Moments
       in a 15\,$M_\odot$ Core-Collapse Supernova Sequence}
\titlerunning{Rotation Imprint on Time-Integrated CCSN Neutrino Spectral Moments}

\author{N.~Viaux\inst{1,2}}
\authorrunning{N. Viaux}

\institute{Departamento de F\'isica, Universidad T\'ecnica Federico Santa Mar\'ia,
           Valpara\'iso, Chile
\and
Millennium Institute for Subatomic Physics at the High Energy Frontier (SAPHIR),
Santiago, Chile}

\abstract{
We study the early post-bounce neutrino signal of the published Garching
$15\,M_\odot$ rotating core-collapse supernova (CCSN) sequence consisting of
non-rotating (NR), slowly rotating (SR, $\Omega_0=0.5$ rad\,s$^{-1}$), and fast-rotating
(FR, $\Omega_0=150$ rad\,s$^{-1}$, artificially boosted ${\sim}300{\times}$) three-dimensional
models. We present a new analysis of these publicly available simulation data; no new
simulations were performed. Our central result, for this specific model sequence, is that SR and FR shift the integrated
spectral moments in \emph{opposite directions} relative to NR: FR drives the spectra
toward softer, more-pinched states, while SR moves them weakly toward harder, less-pinched
states. Placed in a spectral-shift plane $(\Delta\langle E\rangle_L,\,\Delta\alpha_L)$, NR
sits at the origin, and SR and FR occupy \emph{diagonally opposite quadrants}, making the
non-monotonic response immediately visible as an anti-correlation in two spectral
dimensions simultaneously. The focus is the accretion interval
$t_{\rm pb}=0.05$--$0.30$\,s, where the rotation imprint is strongest. Quantitatively,
fast rotation produces $\Delta\langle E_{\nu_e}\rangle_L=-0.513$\,MeV and
$\Delta\alpha_{\nu_e}=+0.161$, with corresponding shifts
$\Delta\langle E_{\bar\nu_e}\rangle_L=-0.440$\,MeV and $\Delta\alpha_{\bar\nu_e}=+0.173$;
the SR shifts are an order of magnitude smaller and in the opposite sense.
The fast-rotation signature is coherent across all $15\,488$ lines of sight and is
established during early accretion. With only three models from a single progenitor
family, this result is a phenomenological characterization of one published sequence
and a suggestive indication of a non-monotonic, possibly strongly nonlinear,
rotational response within this sequence; the functional form and generality of the
dependence on $\Omega_0$ remain unconstrained.
Fast rotation widens the integrated flavor hierarchies while leaving the
net electron-lepton-number excess nearly unchanged ($0.3\%$), indicating spectral
reorganization rather than gross suppression. Under strongly idealized assumptions,
the FR$-$NR spectral difference maps to non-negligible model-to-model event-count
differences in Hyper-Kamiokande, DUNE, and JUNO at Galactic distances; these represent
conditional model-separability estimates within this one sequence, not observational
detection significances of rotation.
}


\maketitle

\section{Introduction}

Rotation is one of the key physical ingredients that can alter the outcome and
observational signature of a core-collapse supernova (CCSN). Even modest pre-collapse
rotation is amplified by several orders of magnitude during stellar core contraction,
and can reach dynamically important levels in the iron core at the onset of collapse
\cite{Heger2005,Ott2006}. Once the core begins to rotate rapidly, centrifugal forces
modify the structure of the proto-neutron star (PNS) and its surrounding accretion flow,
alter the neutrino-driven gain region, and change the character of
hydrodynamic instabilities such as the standing accretion-shock instability (SASI)
\cite{Ott2006,Blondin2007,Iwakami2009}. The resulting perturbations are potentially
detectable in both gravitational waves \cite{Dimmelmeier2008,Abdikamalov2014,Richers2017}
and neutrinos, making the rotation imprint in the neutrino signal a central question for
multi-messenger CCSN astronomy.

Neutrino signals from rotating CCSNe have been studied analytically and numerically from
several complementary perspectives. High-resolution axisymmetric (2D) models have shown
that rapid rotation produces strong centrifugal bounce signals and modifies the early
post-bounce accretion phase \cite{Ott2006,Dimmelmeier2008}. Three-dimensional models,
now reaching sufficient resolution to capture both rotation and turbulent fluid
instabilities self-consistently, reveal that even modest rotation can trigger a spiral
SASI mode and introduce strong directional asymmetries in the neutrino emission
\cite{Blondin2007,Walk2018,Walk2019,Powell2020}. These angle-dependent, time-domain
signatures have been proposed as diagnostics for rotation identification in real-time
burst data from running or planned neutrino detectors
\cite{Walk2018,Walk2019,Tamborra2014b}.

For the Garching $15\,M_\odot$ rotating sequence studied here, previous work has
established several important angle-dependent signatures. Walk et al.\ \cite{Walk2018}
showed that SASI-driven modulation creates a distinctive time-frequency fingerprint
in the event-rate signal that depends strongly on the observer direction, and that
a differential analysis of signals from multiple detectors can act as a rotation
``gyroscope.'' Walk et al.\ \cite{Walk2019} subsequently analyzed the lepton-emission
self-sustained asymmetry (LESA) \cite{Tamborra2014} and the angular structure of the
neutrino transport in the same model family, finding that fast rotation substantially
modifies the LESA-driven flux asymmetry. These results establish the Garching
NR/SR/FR sequence as the most extensively characterized set of published
three-dimensional rotating CCSN models.

The question addressed in the present paper is complementary to those earlier studies.
Whereas prior analyses focused on time-domain modulation amplitudes, SASI mode
identification, and angle-dependent event-rate oscillations, we ask a different
question: if one \emph{integrates} over the full early-accretion epoch
$t_{\rm pb}=0.05$--$0.30$\,s and projects the signal onto low-dimensional spectral
observables --- the luminosity-weighted mean energy $\langle E\rangle_L$ and the
spectral-shape (pinch) parameter $\alpha$ --- does rotation still leave a distinct
imprint? To our knowledge, the time-integrated $\langle E\rangle_L$ and $\alpha_L$
for this specific Garching rotating sequence have not been reported in the literature;
the prior analyses focused on time-resolved or direction-resolved signatures rather
than these integrated moments. This is a meaningful question because real neutrino
telescopes measure time-integrated or binned quantities over finite detector windows,
and the diagnostic value of the signal depends on how much information survives the
time averaging inherent in finite-duration detection. The
answer turns out to be affirmative and carries a non-obvious structure: the slow and
fast models do not simply sample a smooth continuum of rotation-induced changes, but
instead the spectral response is nearly flat between NR and SR and then jumps
qualitatively in FR. Within this model sequence the spectral response is non-monotonic: SR and FR
occupy opposite quadrants of the spectral-shift plane, and the SR shift is an
order of magnitude smaller than FR. We acknowledge that three models are insufficient
to distinguish a true discontinuous threshold from a steep power-law or other
nonlinear dependence on $\Omega_0$.

This non-monotonic response has direct observational consequences. The next generation of
neutrino detectors --- Hyper-Kamiokande \cite{HyperK2018}, DUNE \cite{DUNE2020},
and JUNO \cite{JUNO2016} --- will collect thousands of events from a Galactic CCSN in
the first few hundred milliseconds after bounce. The spectral mean-energy and pinch-parameter
shifts we quantify translate into model-to-model event-count differences that we use
to illustrate the conditional model separability under idealized detector assumptions
in Section~\ref{sec:detection}.

The contribution of this paper is therefore: (i) a \emph{new analysis} of publicly
available 3D simulation data (no new simulations were performed); (ii) the
identification of a \emph{compact phenomenological signature} — the anti-correlated
displacement in the $(\langle E\rangle_L,\alpha_L)$ plane — that is not visible in
the time-domain or calorimetric observables emphasized by prior work on the same
sequence; and (iii) an exploration of its possible observational utility under
idealized detector assumptions. Time-integrated analyses of this kind are less
sensitive to the specific time profile of the emission but provide a single-number
summary that can be compared across model families and evaluated with modest
detector statistics. We stress that all conclusions in this paper are
drawn from a \emph{single} progenitor mass, equation of state, and transport
framework, with the FR model using an artificially boosted rotation rate; the
generality of the findings across progenitor structure, EOS, and rotation profile
remains to be established with a broader set of 3D rotating models.

The paper is organized as follows. Section~\ref{sec:models} describes the three-dimensional
Garching models and the spectral observables used. Section~\ref{sec:results} presents the
main results: spectral planes, time evolution, integrated response, and directional
(skymap) robustness. Section~\ref{sec:discussion} provides the physical interpretation.
Section~\ref{sec:detection} provides an illustrative model-separability analysis under idealized detector assumptions. Section~\ref{sec:conclusions}
draws the conclusions.

\section{Models and Observables}
\label{sec:models}

\subsection{The Garching $15\,M_\odot$ rotating sequence}

We use the published Garching three-dimensional $15\,M_\odot$ model sequence based on
the Heger et al.\ $m15u6$ progenitor \cite{Heger2005}: NR (non-rotating),
SR ($\Omega_0=0.5$\,rad\,s$^{-1}$), and FR ($\Omega_0=150$\,rad\,s$^{-1}$, artificially
boosted ${\sim}300{\times}$ above the self-consistent stellar-evolution value of
${\sim}0.007$\,rad\,s$^{-1}$ at the iron-core centre)
\cite{Summa2018,Walk2018,Walk2019}. The SR model uses the self-consistent differential
rotation profile from the stellar evolution calculation, while the FR model employs an
artificially enhanced central angular velocity, in order to explore the dynamically
important fast-rotation regime. For context, $\Omega_0=150$\,rad\,s$^{-1}$ corresponds
to an initial spin period of $P_0 \approx 42$\,ms, comparable to the most rapidly
rotating magnetar progenitors inferred from observations of aspherical supernova
remnants and magnetar birth statistics, where rotation periods as short as a few
milliseconds are expected. The FR model therefore represents an extreme but not
physically impossible pre-collapse rotation state. The neutrino transport is performed with the VERTEX/VEF code using a variable
Eddington factor computed from a model Boltzmann equation \cite{Rampp2002}; all three
models use the LS220 equation of state \cite{Lattimer1991LS220}.

The neutrino properties are extracted at $r=500$\,km on a full spherical grid consisting
of $N_\theta = 88$ polar angles and $N_\phi = 176$ azimuthal angles, giving
$N_{\rm ang} = 15\,488$ lines of sight with solid-angle weights
$\Delta\Omega_{ij} = \sin\theta_i\,\Delta\theta\,\Delta\phi$ (summing to $4\pi$). For each direction $(\theta_i,\phi_j)$ the files provide the observer-projected
\emph{equivalent isotropic luminosity} $L^{\rm obs}(\theta,\phi;t)$ --- the
luminosity that a distant observer in direction $(\theta,\phi)$ would infer assuming
the source radiates isotropically, i.e., $L^{\rm obs} = 4\pi r^2 F_\nu$ where
$F_\nu$ is the local energy flux --- together with the luminosity-weighted mean energy
$\langle E\rangle^{\rm obs}(\theta,\phi;t)$, and the luminosity-weighted mean squared
energy $\langle E^2\rangle^{\rm obs}(\theta,\phi;t)$, separately for $\nu_e$,
$\bar\nu_e$, and $\nu_x$ (representing a single heavy-lepton species). The all-sky averages are formed as solid-angle-weighted sums
$\langle Q\rangle = \sum_{ij} L^{\rm obs}_{ij}\,Q_{ij}\,\Delta\Omega_{ij} /
\sum_{ij} L^{\rm obs}_{ij}\,\Delta\Omega_{ij}$, i.e., luminosity-weighted (flux-weighted)
over the angular grid; the solid-angle weights $\Delta\Omega_{ij}$ ensure uniform
spherical coverage.

\subsection{Spectral observables}

The analysis window is fixed to $t_{\rm pb}=0.05$--$0.30$\,s. This choice is motivated
as follows: the lower bound $t_{\rm pb} = 0.05$\,s excludes the prompt
electron-neutrino neutronization burst (which peaks near $t_{\rm pb} \simeq 0.005$\,s
and is not the focus of this study), while the upper bound $t_{\rm pb} = 0.30$\,s
corresponds to the last available time step common to all three models, retaining the
full early accretion phase without extrapolation. The FR rotation imprint is already
detectable by $t_{\rm pb} \simeq 0.07$\,s, so the choice of upper bound does not
strongly affect the FR$-$NR significance.

For each flavor we compute:
\begin{itemize}
\item The total emitted energy $E_\nu^{\rm tot} = \int L_\nu(t)\,dt$, integrated over the window.
\item The peak luminosity $L_\nu^{\rm pk} = \max_t L_\nu(t)$.
\item The luminosity-weighted mean energy
      \begin{equation}
        \langle E_\nu\rangle_L = \frac{\int L_\nu(t)\,\langle E_\nu\rangle(t)\,dt}
                                      {\int L_\nu(t)\,dt}.
      \end{equation}
\item The integrated pinch parameter, defined from the luminosity-weighted moments as
      \begin{equation}
        \alpha_\nu = \frac{2\langle E\rangle_L^2 - \langle E^2\rangle_L}
                         {\langle E^2\rangle_L - \langle E\rangle_L^2},
        \label{eq:alpha}
      \end{equation}
      where $\langle E^2\rangle_L = \int L\langle E^2\rangle dt / \int L\,dt$.
\end{itemize}
The parameter $\alpha$ measures spectral pinching: larger $\alpha$ means a narrower,
more quasi-thermal spectrum. For a Gamma-distribution spectrum,
$\alpha = (\langle E\rangle^2/\sigma_E^2) - 2$ where $\sigma_E^2 = \langle E^2\rangle - \langle E\rangle^2$.

We also monitor the integrated flavor-hierarchy gaps
$\Delta E_{\bar e e} \equiv \langle E_{\bar\nu_e}\rangle_L - \langle E_{\nu_e}\rangle_L$
and $\Delta E_{xe} \equiv \langle E_{\nu_x}\rangle_L - \langle E_{\nu_e}\rangle_L$,
and the net emitted electron-lepton-number excess
$N_{\rm ELN} \propto N_{\nu_e} - N_{\bar\nu_e}$, where
$N_\nu \propto E_\nu^{\rm tot}/\langle E_\nu\rangle_L$.

We stress that the integrated pinch parameter defined in Eq.~(\ref{eq:alpha}) is
\emph{not} the arithmetic mean of $\alpha(t)$ over the window, but rather the
spectral shape parameter inferred from the time-integrated moments. The two
definitions agree when the spectral shape does not evolve over the window, but
diverge when $\alpha(t)$ varies significantly. Since the spectral shape in the
accretion phase does evolve (the spectra harden as the luminosity declines),
the integrated $\alpha_L$ carries information about the time-evolution of the
spectral shape that the mean of instantaneous $\alpha(t)$ would dilute. Our
definition gives a cleaner physical interpretation: $\alpha_L$ measures the
effective spectral shape of the \emph{cumulative} neutrino dose received by a
detector over the window, and directly enters the detectability calculations
in Section~\ref{sec:detection}. For the NR model, the luminosity-weighted mean of
the instantaneous $\alpha_{\bar\nu_e}(t)$ over the window differs from $\alpha_L$ by
approximately $5$--$8\%$ (the integrated value is slightly higher because the early,
high-luminosity portion of the burst has a lower pinch than the late accretion phase
at $t_{\rm pb} > 0.20$\,s, and the luminosity-weighted average in Eq.~\ref{eq:alpha}
up-weights that early phase).

The complete set of integrated quantities for all three models is collected in
Table~\ref{tab:integrated} (Appendix~\ref{app:tables}).

\section{Results}
\label{sec:results}

\subsection{Integrated spectral observables}

\begin{figure*}[t]
\centering
\includegraphics[width=\textwidth]{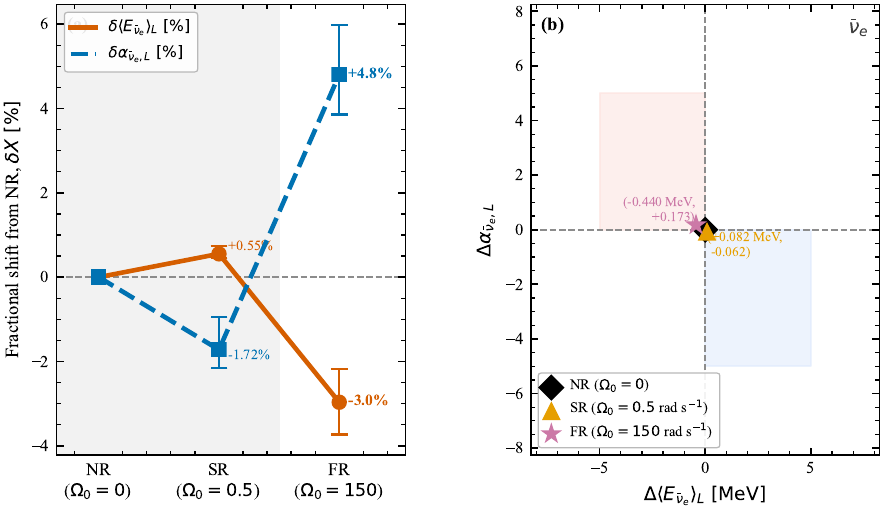}
\caption{Central result: non-monotonic, opposite-direction spectral response
  to rotation, integrated over $t_{\rm pb}=0.05$--$0.30$\,s.
  \textbf{(a)} Fractional shifts $\delta\langle E_{\bar\nu_e}\rangle_L$
  (orange-red circles, Eq.~\ref{eq:fracshift}) and $\delta\alpha_{\bar\nu_e,L}$
  (blue squares) plotted along the rotation sequence NR$\to$SR$\to$FR.
  The two curves open as a scissors: SR moves slightly in one direction while
  FR moves strongly in the \emph{opposite} direction in both spectral dimensions
  simultaneously, demonstrating a non-monotonic response to increasing rotation rate.
  The 68\% sky-percentile intervals (16th--84th percentile of the $15\,488$ per-direction
  values) indicate the directional spread of the emission, \emph{not} measurement
  uncertainty.
  \textbf{(b)} Per-direction scatter in the
  $(\Delta\langle E_{\bar\nu_e}\rangle_L,\,\Delta\alpha_{\bar\nu_e,L})$ plane.
  Each point is one of the $15\,488$ observer directions.
  NR (grey) clusters near the origin; SR (orange) is displaced slightly in one
  quadrant; FR (magenta) is displaced strongly in the diagonally opposite quadrant.
  The non-overlap of the FR and NR clouds confirms that the spectral imprint is
  coherent across all viewing angles.}
\label{fig:main}
\end{figure*}

A direct way to expose the rotation-induced spectral change is through
the fractional shift of an integrated observable $X \in \{\langle E\rangle_L,\,\alpha_L\}$
relative to the non-rotating reference,
\begin{equation}
\delta X \equiv \frac{X_{\rm model} - X_{\rm NR}}{X_{\rm NR}} \times 100\%\,.
\label{eq:fracshift}
\end{equation}
Plotting $\delta\langle E_{\bar\nu_e}\rangle_L$ and $\delta\alpha_{\bar\nu_e,L}$
along the rotation sequence NR\,$\to$\,SR\,$\to$\,FR immediately exposes the
central non-monotonic behavior of this paper, summarized in Fig.~\ref{fig:main}.

\paragraph{Panel~(a) — the scissors diagram.}
The two curves traced in panel~(a) — $\delta\langle E_{\bar\nu_e}\rangle_L$ (orange-red, solid) and
$\delta\alpha_{\bar\nu_e,L}$ (blue, dashed) — open like a pair of scissors in \emph{opposite directions}
along NR\,$\to$\,SR\,$\to$\,FR:

\begin{itemize}
\item From NR to SR the mean energy rises slightly ($\delta\langle E_{\bar\nu_e}\rangle_L
  = +0.55\%$, $+0.082$\,MeV) while the pinch parameter falls ($\delta\alpha_{\bar\nu_e}
  = -1.7\%$, $\Delta\alpha=-0.062$). The spectrum is marginally harder and
  less pinched than the non-rotating baseline.

\item From NR to FR both curves reverse sign and jump to much larger amplitudes:
  the mean energy decreases sharply ($\delta\langle E_{\bar\nu_e}\rangle_L = -3.0\%$,
  $-0.440$\,MeV) and the pinch parameter increases ($\delta\alpha_{\bar\nu_e} = +4.8\%$,
  $\Delta\alpha = +0.173$). The spectrum is appreciably softer and more pinched.

\item The SR and FR values therefore have \emph{opposite signs} in both $\delta\langle
  E\rangle_L$ and $\delta\alpha_L$ simultaneously. No simple linear interpolation
  connects NR, SR, and FR: the spectral response is non-monotonic.
\end{itemize}

The same anti-correlated pattern holds for $\nu_e$, with slightly larger amplitudes
($\delta\langle E_{\nu_e}\rangle_L = -4.3\%$ and $\delta\alpha_{\nu_e} = +5.5\%$ for FR),
confirming that the effect is not specific to a single flavor channel.
The directional spread indicated by the 16th--84th percentile interval over the
$15\,488$ per-direction values (a measure of spatial anisotropy, not measurement
uncertainty) shows that the FR shift is robustly resolved against directional scatter,
while the SR shift is consistent with zero within its spread. The all-sky mean values are listed
in Table~\ref{tab:integrated}.

\paragraph{Note on the scissors shape.}
The anti-correlated shift in $\langle E\rangle_L$ and $\alpha_L$ is phenomenologically
coherent: a softer spectrum and a higher pinch parameter both result from emission
dominated by lower-temperature conditions. One plausible interpretation, detailed further
in Sect.~\ref{sec:discussion}, is that fast rotation centrifugally modifies the accretion
geometry and shifts the effective neutrinosphere to larger, cooler radii in the
equatorial belt, yielding the observed softer and more pinched integrated spectra. The
scissors shape is the spectral representation of this non-monotonic response; its
physical origin is discussed in Sect.~\ref{sec:discussion}.

\paragraph{Panel~(b) — per-direction scatter.}
Panel~(b) shows the same information in the two-dimensional
$(\Delta\langle E_{\bar\nu_e}\rangle_L,\,\Delta\alpha_{\bar\nu_e,L})$ shift plane,
with each of the $15\,488$ observer directions plotted as a translucent point.
NR (grey) forms a compact cloud centred at the origin; its finite spread reflects
the directional anisotropy intrinsic to the non-rotating model. SR (orange) is
displaced only slightly from NR, in the harder/less-pinched direction.
FR (magenta) is displaced far from the origin in the diagonally opposite
(softer/more-pinched) direction, and its cloud does not overlap the NR cloud —
confirming that the fast-rotation spectral imprint is significant and coherent
for every observer direction, not merely on average. The full angular analysis
via sky maps is presented in Section~\ref{sec:skymap}.

\subsection{Time evolution}

\begin{figure}[t]
\centering
\includegraphics[width=\columnwidth]{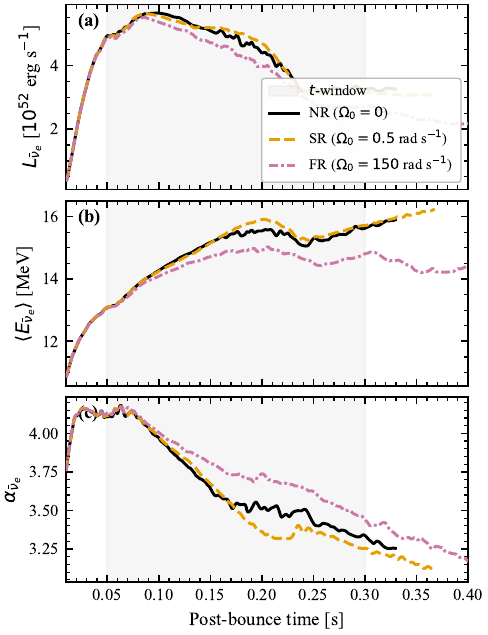}
\caption{All-sky-averaged time evolution of the $\bar\nu_e$ signal for the three
  Garching models.
  \textbf{(a)} Luminosity $L_{\bar\nu_e}(t)$.
  \textbf{(b)} Mean energy $\langle E_{\bar\nu_e}\rangle(t)$.
  \textbf{(c)} Pinch parameter $\alpha_{\bar\nu_e}(t)$.
  Colors: NR = black solid, SR = orange dashed, FR = magenta dash-dot.
  The shaded strip marks the integration window $0.05$--$0.30$\,s.
  The spectral quantities (b) and (c) show a clear and persistent separation between
  FR and the NR/SR pair, while the luminosity differences are comparatively small.}
\label{fig:time}
\end{figure}

Understanding \emph{when} the rotation signature builds up is as important as its
amplitude. The all-sky-averaged time series (Fig.~\ref{fig:time}) shows that the
luminosity curves are broadly similar across all three models — the total energy
release is only weakly sensitive to the accretion geometry — whereas the spectral
channels diverge early and persistently.

The FR model develops a systematically softer $\langle E_{\bar\nu_e}\rangle(t)$ and
more pinched $\alpha_{\bar\nu_e}(t)$ that is already visible by
$t_{\rm pb}\simeq 0.07$\,s and stabilizes by $\simeq 0.12$\,s, well within the
early-accretion window. The early onset reflects the rapid dynamical response of
the accretion flow: once the infalling material carries enough angular momentum to
significantly alter the accretion geometry near the PNS surface, the neutrinosphere
settles into the rotationally modified configuration on the
free-fall timescale ($\sim$10\,ms). The SR model, by contrast, tracks NR throughout
the entire window: the instantaneous $|\langle E_{\bar\nu_e}\rangle_{\rm SR}(t)
- \langle E_{\bar\nu_e}\rangle_{\rm NR}(t)|$ never exceeds $0.15$\,MeV ($< 1\%$)
at any given time over $t_{\rm pb} = 0.05$--$0.30$\,s, compared to the FR departure of up to $0.6$\,MeV.
It does not represent a scaled-down version of the FR trend but
an evolution that is quantitatively close to the non-rotating case. This is a key
diagnostic: if the rotation response were simply linear in $\Omega_0$, SR would sit
midway between NR and FR at all times. The fact that it does not is the clearest
time-domain signature of this non-monotonic response.

How the integrated shift builds up as a function of the integration window is
shown in Fig.~\ref{fig:summary}(c). The FR$-$NR mean-energy difference accumulates
rapidly, reaching $2\%$ by $t_{\rm max}\simeq 0.12$\,s and its maximum near $0.30$\,s.
Including later emission ($t > 0.30$\,s) dilutes rather than amplifies the signal:
as the PNS cooling phase begins to dominate, the spectral divergence between models
narrows because the long-term cooling trajectory is less sensitive to the early
accretion geometry. The integration window $0.05$--$0.30$\,s chosen here is
therefore near-optimal for isolating the rotation imprint.

\subsection{Integrated observable response}

\begin{figure*}[t]
\centering
\includegraphics[width=\textwidth]{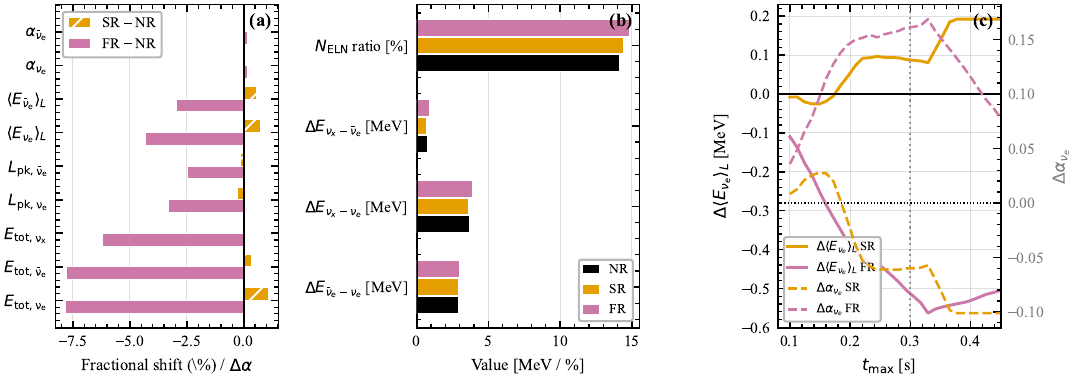}
\caption{Integrated response of the rotating sequence.
  \textbf{(a)} Fractional shifts of positive observables relative to NR for both SR
  (orange) and FR (magenta). The largest shifts are in the spectral channels; luminosity
  and calorimetric observables respond more weakly.
  \textbf{(b)} Absolute values and model differences for the flavor-hierarchy gaps
  $\Delta E_{\bar\nu_e-\nu_e}$, $\Delta E_{\nu_x-\nu_e}$,
  $\Delta E_{\nu_x-\bar\nu_e}$, and the normalized electron-lepton-number excess.
  \textbf{(c)} Dependence of the FR$-$NR and SR$-$NR integrated mean-energy and pinch
  shifts on the upper integration limit $t_{\rm max}$, demonstrating that the
  fast-rotation signature is established during early accretion and persists robustly.}
\label{fig:summary}
\end{figure*}

Extending the analysis beyond the $\bar\nu_e$ mean energy and pinch parameter
to the full set of tracked observables (Fig.~\ref{fig:summary}) deepens the
physical picture and confirms the non-monotonic hierarchy.
Panel~(a) compares fractional shifts of all tracked observables relative to NR.
The largest fractional changes in FR are in the spectral channels: $-4.3\%$ in
$\langle E_{\nu_e}\rangle_L$, $-3.0\%$ in $\langle E_{\bar\nu_e}\rangle_L$,
$+5.5\%$ in $\alpha_{\nu_e}$, and $+4.8\%$ in $\alpha_{\bar\nu_e}$. By comparison,
the total emitted energies decrease by $\simeq 7.8\%$ (electron flavors) and the peak
luminosities by $2.5$--$3.3\%$, while $\nu_x$ shows smaller changes throughout.
SR shifts are consistently small and in the opposite direction to FR in the spectral
quantities.

Panel~(b) isolates the flavor-hierarchy and lepton-number diagnostics. FR widens the
integrated energy hierarchy: $\Delta E_{\bar e e}$ grows from $2.937$ to $3.010$\,MeV
(+2.5\%), and $\Delta E_{x e}$ grows from $3.673$ to $3.872$\,MeV (+5.4\%). Meanwhile,
the net electron-lepton-number excess $N_{\rm ELN}$ changes by only $0.3\%$. We note
that this near-invariance may be partly accidental due to a partial cancellation between
the decreased $\nu_e$ total energy and the shift in $\langle E_{\nu_e}\rangle_L$, and
should not be interpreted as evidence that rotation fundamentally preserves the lepton
balance. With this caveat, the small change shows that fast rotation primarily
redistributes the spectral structure across flavors rather than simply suppressing
the total electron-flavor lepton output. The SR model again
behaves differently: it slightly raises $N_{\rm ELN}$ and narrows the energy hierarchy,
reinforcing the non-monotonic picture.

Panel~(c) shows the window-scan dependence. Defining the build-up time as the
$t_{\rm max}$ at which the FR$-$NR mean-energy shift first exceeds $2\%$, this occurs
at $t_{\rm max} \simeq 0.10$\,s for $\nu_e$ and $\simeq 0.12$\,s for $\bar\nu_e$.
The pinch-parameter difference crosses the $5\%$ level at $t_{\rm max} \simeq 0.15$\,s.
The imprint is strongest in the window $0.05$--$0.30$\,s chosen here and is then
diluted as later emission is included. The SR$-$NR shift remains consistent with zero
over the entire scanned range $0.10$--$0.45$\,s, confirming that the non-monotonic
behavior is not a transient feature of a particular time interval but a systematic
property of the early-accretion phase.

\subsection{Directional structure: all-sky temporal variability maps}
\label{sec:skymap}

The Garching OBS files are angle-resolved on a full $88\times 176$ angular grid,
giving 15\,488 independent lines of sight. Rather than time-integrating the spectral
observables (which washes out transient anisotropies), we map the \emph{temporal
standard deviation} of the instantaneous $\bar\nu_e$ spectral quantities per direction,
\begin{equation}
  \sigma[\langle E_{\bar\nu_e}\rangle_L^{\rm LOS}](\hat{n}) \;\equiv\;
  \mathrm{std}_{t\in[0.05,0.30]\,\rm s}\!\left[\langle E_{\bar\nu_e}\rangle_L(t,\hat{n})\right],
\end{equation}
and equivalently for the pinch parameter $\alpha_{\bar\nu_e,L}$.
This quantity measures how strongly the neutrino spectrum fluctuates in time
as seen from a given observer direction, and is directly sensitive to the large-scale
hydrodynamic instabilities (SASI, LESA, rotation-driven convection) that dominate
the post-bounce accretion phase.

\begin{figure*}[t]
\centering
\includegraphics[width=\textwidth]{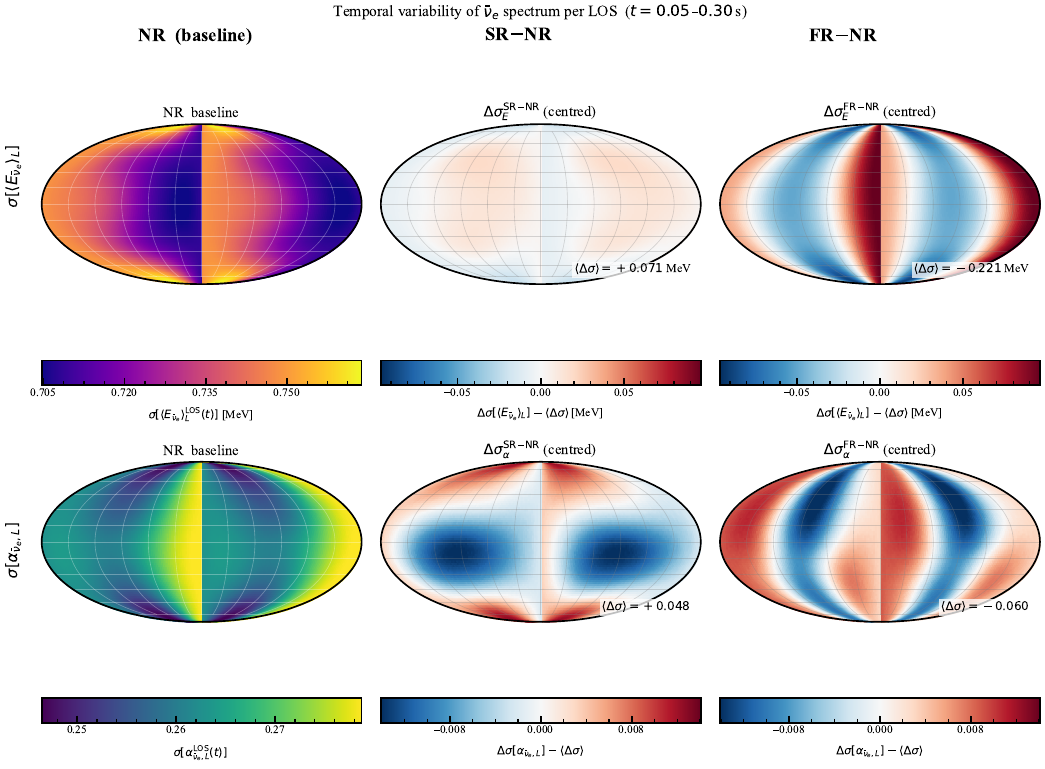}
\caption{Mollweide projection sky maps of the \emph{temporal standard deviation}
  of the $\bar\nu_e$ spectral observables for the Garching $15\,M_\odot$ sequence,
  computed over $t_{\rm pb}=0.05$--$0.30$\,s across all 15\,488 lines of sight.
  \textbf{Left column}: NR baseline maps of $\sigma[\langle E_{\bar\nu_e}\rangle_L^{\rm LOS}(t)]$
  (top, \texttt{plasma} scale) and $\sigma[\alpha_{\bar\nu_e,L}^{\rm LOS}(t)]$ (bottom,
  \texttt{viridis} scale). Spherical-harmonic decomposition (see text) shows these maps
  are isotropic at $\ell\leq 2$; the apparent hemisphere contrast is not confirmed as a
  significant $\ell=1$ feature.
  \textbf{Centre column}: sky-mean-subtracted residuals $\Delta\sigma - \langle\Delta\sigma\rangle_{\rm sky}$
  for SR$-$NR; the annotated $\langle\Delta\sigma\rangle$ gives the removed isotropic offset.
  Weak $\ell=2$ quadrupole structure is present (see text); no dominant mode.
  \textbf{Right column}: same for FR$-$NR. The energy residual is dominated by an
  $\ell=2$ quadrupole carrying ${\sim}74\%$ of map power, with dominant mode
  $(\ell=2,\,m=\pm 2)$ (see text). In the Garching model coordinate system the FR
  rotation axis is aligned with the $z$-axis, corresponding to the map poles at
  Galactic latitude $b=\pm 90^\circ$.
  All maps use Galactic coordinates ($l,b$); north is up, east is left. The
  directional spread intervals (16th--84th sky percentile) shown in
  Fig.~\ref{fig:main} are a measure of spatial anisotropy, not measurement uncertainty.}
\label{fig:skymap}
\end{figure*}

The left column of Fig.~\ref{fig:skymap} shows the NR baseline maps;
the centre and right columns show the sky-mean-subtracted residuals
$\Delta\sigma - \langle\Delta\sigma\rangle_{\rm sky}$ for SR$-$NR and FR$-$NR,
with the removed isotropic offset annotated in each panel. Three distinct physical
regimes emerge.

We present these maps as \emph{exploratory} visualizations, secondary to the main
integrated-spectral result, and quantify their angular structure via a real
spherical-harmonic (SH) decomposition through $\ell=2$. We compute the per-mode
power amplitude $\sqrt{C_\ell}$, where $C_\ell = (2\ell+1)^{-1}\sum_m a_{\ell m}^2$
and $a_{\ell m} = \int f\,Y_{\ell m}\,d\Omega$.

\textit{NR baseline.} Both $\sigma_E$ and $\sigma_\alpha$ maps are monopole-dominated
and isotropic at $\ell\leq 2$: the dipole amplitude is $\sqrt{C_1}\simeq 0.002$\,MeV
(${\sim}0.2\%$ of the map RMS), and the quadrupole $\sqrt{C_2}\simeq 0.020$\,MeV
(${\sim}2.8\%$ of RMS). The apparent hemisphere contrast visible in the color map is
therefore not confirmed as a significant $\ell=1$ feature by the SH decomposition;
any connection to LESA remains speculative at this level of analysis.

\textit{SR$-$NR centred residuals.} Slow rotation amplifies the sky-mean temporal
variance uniformly ($\langle\Delta\sigma_E\rangle = +0.071$\,MeV,
$\langle\Delta\sigma_\alpha\rangle = +0.048$).
After subtracting this isotropic offset, the SH decomposition gives
$\sqrt{C_2}\simeq 2\,\sqrt{C_1}$ for both the energy and pinch maps, with no
dominant angular mode. The quadrupole accounts for $\sim\!80\%$ of
the (dipole+quadrupole) power in the energy residual and $\sim\!91\%$ in the pinch
residual, but neither map is strongly anisotropic at $\ell\leq 2$.

\textit{FR$-$NR centred residuals.} Fast rotation strongly suppresses overall
temporal fluctuations ($\langle\Delta\sigma_E\rangle = -0.221$\,MeV,
$\langle\Delta\sigma_\alpha\rangle = -0.060$), consistent with centrifugal quenching
of SASI \citep{Walk2018}. The SH decomposition of the centred energy residual shows
a dominant $\ell=2$ quadrupole: $\sqrt{C_2}=0.066$\,MeV carrying $\sim\!74\%$ of the
total map power, compared with $\sqrt{C_1}=0.006$\,MeV. The dominant mode is
$(\ell=2,\,m=\pm 2)$, an azimuthal four-lobe pattern, rather than the
$(\ell=2,\,m=0)$ polar-equatorial contrast one might expect from simple spin-axis
imprinting. The $\alpha$ residual is similarly quadrupole-dominated
($\sqrt{C_2}=0.009$\,MeV, ${\sim}95\%$ of dipole+quadrupole power).
These results confirm significant $\ell=2$ angular structure in the FR$-$NR residuals,
while the specific physical origin of the dominant $(m=\pm 2)$ mode requires
further investigation.

\section{Physical Interpretation}
\label{sec:discussion}

\subsection{Non-monotonic spectral response and its physical interpretation}

The non-monotonic spectral sequence is suggestive of a strongly nonlinear dependence
of the accretion geometry on rotation rate, though the available three models cannot
rigorously distinguish a threshold from a steep continuous dependence on $\Omega_0$.
The order-of-magnitude argument below is illustrative only. In the SR model ($\Omega_0=0.5$\,rad\,s$^{-1}$),
the specific angular momentum of the infalling material is too small to reorganize the
accretion geometry near the gain region and PNS surface. The system evolves much as the
non-rotating model would, and the integrated spectral observables remain close to NR.
In the FR model ($\Omega_0=150$\,rad\,s$^{-1}$, boosted $\sim 300\times$), centrifugal
support is strong enough to produce a qualitatively different accretion geometry:
the infalling matter forms a rotationally supported ring or torus near the PNS,
modifying both the ram pressure on the PNS surface and the thermodynamic conditions
in the neutrino-decoupling region \cite{Summa2018,Walk2018}.

A plausible spectral consequence, consistent with the simulated behavior, is that
decoupling occurs at lower density and temperature in the rotationally modified
accretion geometry, yielding softer and more narrowly distributed (higher-$\alpha$)
spectra. One possible picture is a rotationally widened density gradient at the
neutrinosphere: centrifugally supported material in the equatorial belt may
reduce the effective radial temperature gradient seen by equatorial observers,
leading to smaller energy variance and hence higher $\alpha$. We emphasize that
this is an interpretation of the spectral data; verifying it would require
neutrinosphere-geometry diagnostics from the simulation outputs. A plausible
reason the luminosity responds less strongly is that the total gravitational
energy release is set by the mass-accretion rate and binding energy of the PNS,
quantities that are modified at second order by equatorial centrifugal support
while remaining nearly unchanged in the polar directions.

A heuristic estimate is consistent with this picture. The specific angular
momentum required to provide centrifugal support at radius $r$ near the PNS surface
($r \sim 50$\,km) is $j_{\rm c} \sim \sqrt{GM_{\rm PNS}\,r} \sim 10^{16}$\,cm$^2$\,s$^{-1}$
for $M_{\rm PNS} \sim 1.4\,M_\odot$. Pre-collapse, the SR progenitor core has
$j \sim r_{\rm core}^2 \,\Omega_0 \sim (2\times 10^8\,{\rm cm})^2 \times 0.5\,{\rm rad\,s}^{-1}
\sim 2\times 10^{16}$\,cm$^2$\,s$^{-1}$, which after compression to neutron-star radius
gives roughly comparable specific angular momentum --- at the boundary of the
centrifugally important regime in this heuristic estimate. The FR
model, with $\Omega_0 = 150$\,rad\,s$^{-1}$ (boosted $\sim 300\times$), has a much
larger specific angular momentum and lies well within the centrifugally important regime
by the same estimate. This order-of-magnitude argument is consistent with the observed
spectral behavior, but should not be read as locating a precise transition: the three
models bracket different spectral behaviors without constraining the functional form
of the dependence on $\Omega_0$.

The near invariance of $N_{\rm ELN}$ ($0.3\%$ change in FR relative to NR) adds
an important physical constraint: fast rotation does not primarily suppress or
enhance the net electron-type lepton flux, but instead redistributes energy and
spectral width across flavors. This is consistent with the electron-lepton-number
excess being set by the neutronization depth in the PNS core \cite{Tamborra2014},
which is less sensitive to the accretion-layer geometry than the neutrinosphere
temperature.

\subsection{Connection to fluid instabilities and SASI}

When fast rotation is present, the SASI transitions from a sloshing-dominated
to a spiral-dominated mode \citep{Walk2018,Walk2019}, imprinting a characteristic
large-scale angular pattern on the instantaneous neutrino emission.
Our integrated-over-time analysis averages over many SASI cycles, so the spectral
imprint we measure is not the oscillation amplitude itself but the \emph{mean-field}
reorganization of the accretion geometry induced by the spiral mode and its
angular-momentum redistribution. The residual directional scatter — visible
as directional spread intervals in Fig.~\ref{fig:main} and as the structured patterns in the
sky maps of Fig.~\ref{fig:skymap} — represents the SASI-driven angular anisotropy
that survives time integration, while the mean displacement between NR and FR
is the clean mean-field spectral shift. The two contributions are therefore separable:
the former encodes the SASI amplitude and geometry, the latter the mean-field rotation imprint.

\subsection{ELN and flavor transformation: a caveat}

The near invariance of the integrated ELN excess in FR ($0.3\%$; Table~\ref{tab:integrated})
reflects an accidental partial cancellation between changes in $E_{\nu_e}^{\rm tot}$
and $\langle E_{\nu_e}\rangle_L$, and says nothing about whether fast flavor
transformation \cite{Zaizen2021,Tamborra2021} is enhanced or suppressed by rotation;
assessing that would require angle-resolved transport data beyond the spectral moments
used here.

\subsection{Comparison with existing rotating models}

Several other groups have published rotating CCSN neutrino signals in two or three
dimensions. Ott et al.\ \cite{Ott2006} presented a survey of 2D models spanning a wide
range of precollapse rotation rates and showed that the bounce and ring-down phases
produce strong rotation-dependent signatures, consistent with a strongly nonlinear
response in which centrifugal effects become important above some rotation rate. Powell \& M\"uller \cite{Powell2020}
found that rapid rotation enhances the gravitational-wave emission and modifies the
explosion geometry in 3D models, with accompanying changes in the neutrino luminosities.
Our results are consistent with those trends: the spectral softening and increased pinch
we observe in FR are physically analogous to the reduced accretion-rate and modified
PNS surface conditions seen in other fast-rotation simulations.

A key distinction of the present analysis is that it focuses on \emph{integrated}
spectral moments rather than time-domain oscillation amplitudes or peak luminosities.
The non-monotonic response is clearest in this low-dimensional integrated projection,
and the quantitative shifts ($\sim 0.5$\,MeV in $\langle E_{\nu_e}\rangle_L$,
$\sim 0.16$ in $\alpha_{\nu_e}$) are resolvable under idealized detector assumptions.
Future rotating 3D models with different progenitors, rotation profiles, and equations
of state will be needed to assess whether the non-monotonic behavior is a general
feature or specific to the $15\,M_\odot$ Heger et al.\ progenitor and LS220 EOS.

\section{Detection Prospects}
\label{sec:detection}

\subsection{Method}

The integrated spectral shifts we have identified are physically clean and coherent, but
their observational relevance depends on whether they can be resolved with realistic
neutrino telescopes. To give an \emph{illustrative} estimate, we translate the model
luminosities and spectra into expected event counts for three detector configurations.
We emphasize at the outset that this section employs several idealizing assumptions ---
no neutrino oscillations, 100\% detection efficiency, no detector response function,
no backgrounds, and Poisson-only statistics --- that a realistic sensitivity study would
need to relax. The results should therefore be read as order-of-magnitude estimates rather
than detector forecasts. The three detector configurations considered are Hyper-Kamiokande
(Hyper-K), DUNE, and JUNO.

For inverse-beta-decay (IBD) dominated detectors (Hyper-K and JUNO), the event count is
\begin{equation}
  N_{\rm IBD} = \int_{t_{\rm min}}^{t_{\rm max}}
    \frac{L_{\bar\nu_e}(t)}{4\pi d^2}
    \frac{N_p}{\langle E_{\bar\nu_e}(t)\rangle}
    \langle\sigma_{\rm IBD}(t)\rangle\,dt,
  \label{eq:ibd}
\end{equation}
where $N_p$ is the number of free protons in the fiducial volume and
$\langle\sigma_{\rm IBD}\rangle = \sigma_0\,\langle(E-\Delta)^2\rangle$ with
$\sigma_0 = 9.52\times 10^{-44}$\,cm$^2$\,MeV$^{-2}$ and $\Delta=1.293$\,MeV
the neutron-proton mass difference \cite{Strumia2003}. The spectral average of $(E-\Delta)^2$ over a Gamma distribution yields
$\langle(E-\Delta)^2\rangle = \langle E^2\rangle_L - 2\Delta\langle E\rangle_L + \Delta^2$.
Rather than performing a full time-resolved integration of Eq.~(\ref{eq:ibd}), we
approximate the time-series integral by evaluating the spectral averages at the
\emph{time-integrated} moments $\langle E\rangle_L$ and $\langle E^2\rangle_L$
tabulated in Table~\ref{tab:integrated}, and multiplying by the total emitted energy
$E_{\bar\nu_e}^{\rm tot}$:
$N_{\rm IBD} \approx [E_{\bar\nu_e}^{\rm tot}/(4\pi d^2\langle E_{\bar\nu_e}\rangle_L)]
\cdot N_p \cdot \langle\sigma_{\rm IBD}\rangle_L$.
This approximation is equivalent to treating the entire burst as having a single
effective spectral shape characterized by the time-integrated moments, with
$\langle\sigma_{\rm IBD}\rangle_L \equiv \sigma_0(\langle E^2\rangle_L -
2\Delta\langle E\rangle_L + \Delta^2)$ evaluated from Table~\ref{tab:integrated}.
The approximation is accurate to better than $5\%$ compared to a direct time-resolved
integration for the NR model. For SR and FR, where the time-varying spectral shape
evolves differently, we do not expect the approximation error to be materially larger:
the FR spectrum stabilizes early ($t_{\rm pb}\lesssim 0.12$\,s) so the
time-integrated moments are an accurate summary of the emission-weighted spectral shape
over the window.

For DUNE we use the $\nu_e$ charged-current channel on ${}^{40}$Ar,
$\nu_e + {}^{40}{\rm Ar} \to e^- + {}^{40}{\rm K}^*$, with the cross-section
parameterization from \cite{Tomas2003}:
\begin{equation}
  \sigma_{\nu_e+{\rm Ar}}(E_\nu) \approx 6.5\times 10^{-42}
  (E_\nu/\mathrm{MeV} - 5.5)^2~{\rm cm}^2,
\end{equation}
which is appropriate for $E_\nu \gtrsim 8$\,MeV \cite{Tomas2003,Scholberg2012}.
The strong energy dependence $(E-E_{\rm thr})^2$ amplifies the effect of the $-4.3\%$
shift in $\langle E_{\nu_e}\rangle_L$ on the detected rate. More recent and detailed
calculations of the $\nu_e$-Ar cross section (see, e.g., the DUNE supernova sensitivity
study \cite{DUNE2020}) differ from this parameterization at the level of tens of percent
in absolute normalization; however, the \emph{fractional} event-count deficit
$\Delta N / N_{\rm NR}$ is insensitive to normalization uncertainties at fixed spectral
shape, so the ${\sim}20\%$ deficit estimate is robust against cross-section choice.

Detector parameters assumed:
Hyper-K: fiducial mass $187$\,kton water Cherenkov \cite{HyperK2018},
$N_p = 1.25\times 10^{34}$ free protons.
DUNE: fiducial mass $40$\,kton liquid argon \cite{DUNE2020},
$N_{\rm Ar} = 6.02\times 10^{32}$ argon nuclei.
JUNO: fiducial mass $20$\,kton liquid scintillator \cite{JUNO2016},
$N_p = 1.47\times 10^{33}$ free protons.

We assume no oscillations, 100\% efficiency, and $d=10$\,kpc. The results
(Table~\ref{tab:detection}) measure \emph{conditional model separability} — how
distinguishable FR is from NR if NR is the correct baseline and all other unknowns
are fixed — not detection significances of rotation in a real observation.

\subsection{Scaling results}

Throughout this subsection we use $\Delta N = N_{\rm NR} - N_{\rm model}$
(positive = rotating model produces fewer events than NR).
The full results are tabulated in Table~\ref{tab:detection}
(Appendix~\ref{app:tables}). In brief, at $d=10$\,kpc and under the
idealizing assumptions stated above, the FR model yields event-count
deficits of $\Delta N_{\rm FR} = 1\,266$ (Hyper-K, IBD),
$1\,558$ (DUNE, $\nu_e$ CC on Ar), and $147$ (JUNO, IBD) relative to NR,
corresponding to fractional reductions of $\sim 12\%$, $\sim 18\%$, and
$\sim 12\%$ respectively. The SR model produces small excesses
($\Delta N_{\rm SR} \approx -137$ in Hyper-K, $-265$ in DUNE), consistent
with its near-NR spectral character. The $(E-E_{\rm thr})^2$ energy
dependence of the $\nu_e$-Ar cross section amplifies the larger
$\nu_e$ mean-energy shift ($-4.3\%$) and is the reason DUNE shows the
largest fractional deficit.

A distance-scaling argument: defining $d_3$ as the distance at which the
FR deficit equals three times the Poisson fluctuation of $N_{\rm NR}$
(with $\Delta N / \sqrt{N} \propto d^{-1}$), one obtains
$d_3 \approx 40$\,kpc (Hyper-K) and $57$\,kpc (DUNE) under the same
idealizing assumptions. These figures are presented purely as a
model-separability scaling estimate, not as a prediction of rotation
detectability in a real observation.

\subsection{Schematic oscillation check}

The estimates above assume no oscillations. As a schematic check, applying
adiabatic MSW mixing in the normal ordering
($P(\bar\nu_e\to\bar\nu_e)\approx 0.70$, with $\nu_x$ denoting one
heavy-lepton species as defined in Sect.~\ref{sec:models}) reduces both
$N_{\rm NR}$ and $N_{\rm FR}$ by $\sim 15\%$ while reducing $\Delta N$ by
only $\sim 10\%$, since both models undergo the same transformation. The
inverted ordering ($P\approx 0.30$) gives a $\sim 25\%$ reduction in
$\Delta N$. In either case the FR$-$NR spectral shape difference is not
erased by the mixing. These estimates are purely illustrative; a
self-consistent oscillation treatment (including collective effects and
fast flavor transformation \cite{Tamborra2021}) is beyond the scope of
this paper.

The SR model is not separable from NR by integrated spectral moments alone
($|\Delta N_{\rm SR}| \lesssim 1.3\,\sqrt{N_{\rm NR}}$; Table~\ref{tab:detection}).

\section{Conclusions}
\label{sec:conclusions}

Using the published Garching $15\,M_\odot$ three-dimensional rotating sequence --- a single
progenitor family with one EOS and one transport framework --- we have found that the
early-accretion ($t_{\rm pb}=0.05$--$0.30$\,s) neutrino signal contains a
rotation-associated spectral signature with the following properties within this model
family:

\begin{enumerate}
\item \textbf{Strongly nonlinear, non-monotonic response.} Fast rotation softens both
  $\nu_e$ and $\bar\nu_e$ mean energies ($-4.3\%$ and $-3.0\%$) and increases
  their pinch parameters ($+5.5\%$ and $+4.8\%$), whereas slow rotation remains
  close to the non-rotating case and even shifts weakly in the \emph{opposite} spectral
  direction. The NR$\rightarrow$SR$\rightarrow$FR trajectory in the
  $(\langle E\rangle_L, \alpha_L)$ plane is non-monotonic and inconsistent with
  a linear response to the initial angular velocity within this model family. We note
  that with only three models it is not possible to distinguish a true discontinuous
  threshold from a steep power-law or other strongly nonlinear dependence on $\Omega_0$.

\item \textbf{Spectral reorganization, not gross lepton suppression.} Fast rotation
  widens the flavor-energy hierarchy ($\Delta E_{\bar e e}$: $+2.5\%$;
  $\Delta E_{xe}$: $+5.4\%$) while leaving the net emitted electron-lepton-number
  excess nearly unchanged ($\Delta N_{\rm ELN} = 0.3\%$). The dominant effect is a
  redistribution of spectral width and mean energy across flavors driven by modified
  thermodynamic conditions in the neutrinosphere layer.

\item \textbf{Full-sky coherence.} The per-direction scatter in
  Fig.~\ref{fig:main}(b) confirms that the FR spectral imprint is coherent across all
  $15\,488$ lines of sight: every observer direction satisfies $\Delta\langle E\rangle_L < 0$
  and $\Delta\alpha_L > 0$ for FR relative to NR, while the SR$-$NR scatter shows only
  noise-level displacement. The directional scatter is fully captured by the 68\% sky-percentile
  intervals in Fig.~\ref{fig:main}(a) and does not reverse the FR displacement from NR
  for any observer direction. Figure~\ref{fig:skymap} (temporal-variability maps, a
  separate quantity) provides exploratory directional context and is discussed in
  Sect.~\ref{sec:skymap}.

\item \textbf{Illustrative model separability under idealized assumptions.}
  At $d=10$\,kpc the FR$-$NR rate deficit is $1\,266$ events in Hyper-Kamiokande (IBD),
  $1\,558$ events in DUNE ($\nu_e$ CC on Ar), and $147$ events in JUNO,
  corresponding to 3$\times$-Poisson distance reaches of 40\,kpc (Hyper-K) and
  57\,kpc (DUNE). These model-to-model differences are large compared to Poisson
  fluctuations under the idealized assumptions adopted (no oscillations, 100\%
  efficiency, Poisson-only statistics); they represent a conditional
  model-separability estimate within this one sequence, not a forecast for identifying
  rotation in a real observed burst.
\end{enumerate}

In summary, the present analysis is a phenomenological re-analysis of one published
model sequence. It does not establish a generic rotational threshold in CCSN neutrino
spectra; it identifies a compact integrated-spectral diagnostic that behaves in a
non-monotonic, strongly sequence-dependent way within the available NR/SR/FR dataset,
and explores whether the resulting model-to-model differences are non-negligible at
the level of idealized detector estimates. The result is hypothesis-generating: it
motivates the question of whether the non-monotonic integrated spectral signature
persists across a broader range of progenitor masses, equations of state, and
self-consistent rotation profiles, a question that can only be resolved by a more
complete set of three-dimensional rotating CCSN simulations.

\begin{acknowledgements}
N.V.\ acknowledges support from the Millennium Institute
for Subatomic Physics at the High Energy Frontier (SAPHIR), ICN2019\_044.
The simulation data used in this work are from the Garching Core-Collapse Supernova
Archive (\url{https://wwwmpa.mpa-garching.mpg.de/ccsnarchive/}); the authors of
the original model papers cited herein are credited for providing these data.
\end{acknowledgements}

\appendix

\section{Data format and averaging procedure}
\label{app:data}

The Garching OBS files store the observer-projected neutrino properties on a uniform
$N_\theta \times N_\phi = 88 \times 176$ angular grid at each timestep, giving
$N_{\rm ang} = 15\,488$ angular entries per time and per flavor. The grid is uniformly
spaced in polar angle $\theta_i = (2i-1)\pi / (2N_\theta)$ for $i = 1, \ldots, N_\theta$
and in azimuthal angle $\phi_j = (-N_\phi + 2j - 1)\pi / N_\phi$ for
$j = 1, \ldots, N_\phi$. The solid-angle weight of element $(i,j)$ is
$\Delta\Omega_{ij} = \sin\theta_i \cdot (\pi/N_\theta) \cdot (2\pi/N_\phi)$, with
$\sum_{ij} \Delta\Omega_{ij} = 4\pi$.

The all-sky-averaged luminosity at time $t$ is computed as
\begin{equation}
  L_\nu(t) = \frac{1}{4\pi} \sum_{i,j} L_\nu^{\rm obs}(\theta_i, \phi_j; t)\,
             \Delta\Omega_{ij},
\end{equation}
where $L_\nu^{\rm obs}(\theta,\phi;t)$ is the equivalent isotropic luminosity
as observed from direction $(\theta,\phi)$. The luminosity-weighted mean energy is
\begin{equation}
  \langle E_\nu(t)\rangle = \frac{\sum_{i,j} L_\nu^{\rm obs}\,\langle E_\nu^{\rm obs}\rangle\,
             \Delta\Omega_{ij}}{\sum_{i,j} L_\nu^{\rm obs}\,\Delta\Omega_{ij}},
\end{equation}
and similarly for $\langle E_\nu^2(t)\rangle$. These all-sky averages are then
time-integrated over the fiducial window $[t_{\rm min}, t_{\rm max}]$ using the
trapezoidal rule on the native timestep grid of each model.

For the per-direction (line-of-sight) analysis used in Figs.~\ref{fig:main}
and~\ref{fig:skymap}, the same time integral is applied separately to each angular
bin $k$:
\begin{align}
  \langle E_\nu\rangle_L^{(k)} &= \frac{\int L_\nu^{\rm obs}(\Omega_k;t)\,
                                \langle E_\nu^{\rm obs}\rangle(\Omega_k;t)\,dt}
                               {\int L_\nu^{\rm obs}(\Omega_k;t)\,dt}, \\
  \alpha_\nu^{(k)} &= \frac{2[\langle E\rangle_L^{(k)}]^2 - \langle E^2\rangle_L^{(k)}}
                           {[\langle E^2\rangle_L^{(k)}] - [\langle E\rangle_L^{(k)}]^2},
\end{align}
yielding a full sky map of spectral observables. The all-sky means in
Table~\ref{tab:integrated} are recovered as the solid-angle-weighted average of these
LOS values, and agree to better than $0.01$\,MeV with the directly computed
angle-averaged time series.

The raw data files contain occasional duplicate timesteps; for such entries we retain
only the first $N_{\rm ang}$ angular rows and discard the remainder, following the
convention used by Walk et al.\ \cite{Walk2018,Walk2019}.

The Garching simulation data used in this paper are publicly available through the
Garching Core-Collapse Supernova Archive maintained by the Max-Planck-Institut f\"ur
Astrophysik. The analysis scripts that read these files and produce Figs.~1--4 and
Tables~B.1--B.2 are available from the author upon reasonable request.

\section{Numerical tables}
\label{app:tables}

\begin{table*}[ht]
\caption{Integrated spectral quantities for the Garching $15\,M_\odot$ sequence,
         $t_{\rm pb} = 0.05$--$0.30$\,s. Energies in MeV, total emitted energies in
         $10^{51}$\,erg ($= 1\,\rm B$).}
\label{tab:integrated}
\centering
\small
\begin{tabular}{lcccccc}
\toprule
 & \multicolumn{2}{c}{NR ($\Omega_0=0$)} & \multicolumn{2}{c}{SR ($\Omega_0=0.5$)} & \multicolumn{2}{c}{FR ($\Omega_0=150$)} \\
\cmidrule(lr){2-3}\cmidrule(lr){4-5}\cmidrule(lr){6-7}
Quantity & $\nu_e$ & $\bar\nu_e$ & $\nu_e$ & $\bar\nu_e$ & $\nu_e$ & $\bar\nu_e$ \\
\midrule
$E_\nu^{\rm tot}$ [B]            & 12.00 & 11.26 & 12.13 & 11.30 & 11.07 & 10.39 \\
$L_\nu^{\rm pk}$ [$10^{52}$ erg\,s$^{-1}$] & 6.2 & 5.7 & 6.2 & 5.6 & 6.0 & 5.5 \\
$\langle E\rangle_L$ [MeV]       & 11.896 & 14.833 & 11.985 & 14.915 & 11.383 & 14.393 \\
$\alpha_L$                       & 2.949  & 3.602  & 2.888  & 3.540  & 3.110  & 3.775  \\
\midrule
 & \multicolumn{2}{c}{$\nu_x$} & \multicolumn{2}{c}{$\nu_x$} & \multicolumn{2}{c}{$\nu_x$} \\
\cmidrule(lr){2-3}\cmidrule(lr){4-5}\cmidrule(lr){6-7}
$E_\nu^{\rm tot}$ [B]            & \multicolumn{2}{c}{7.87} & \multicolumn{2}{c}{7.87} & \multicolumn{2}{c}{7.38} \\
$\langle E\rangle_L$ [MeV]       & \multicolumn{2}{c}{15.569} & \multicolumn{2}{c}{15.622} & \multicolumn{2}{c}{15.255} \\
$\alpha_L$                       & \multicolumn{2}{c}{2.679} & \multicolumn{2}{c}{2.629} & \multicolumn{2}{c}{2.775} \\
\midrule
$\Delta E_{\bar e e}$ [MeV]      & \multicolumn{2}{c}{2.937} & \multicolumn{2}{c}{2.930} & \multicolumn{2}{c}{3.010} \\
$\Delta E_{x e}$ [MeV]           & \multicolumn{2}{c}{3.673} & \multicolumn{2}{c}{3.637} & \multicolumn{2}{c}{3.872} \\
$N_{\rm ELN}$ ratio (\%)         & \multicolumn{2}{c}{13.86} & \multicolumn{2}{c}{13.95} & \multicolumn{2}{c}{13.82} \\
\bottomrule
\end{tabular}
\end{table*}

\begin{table*}[ht]
\caption{Expected event counts in the early-accretion window
         $t_{\rm pb}=0.05$--$0.30$\,s for a source at $d=10$\,kpc
         (under the idealized assumptions of Sect.~\ref{sec:detection}).
         $\Delta N_{\rm SR} = N_{\rm NR}-N_{\rm SR}$ and
         $\Delta N_{\rm FR} = N_{\rm NR}-N_{\rm FR}$ (positive value means the
         rotating model detects fewer events than NR).
         The last column gives the Galactic distance $d_3$ at which the FR deficit
         equals three times the Poisson noise $3\sqrt{N_{\rm NR}}$.}
\label{tab:detection}
\centering
\small
\begin{tabular}{llrrrrrrr}
\toprule
Detector & Channel & $N_{\rm NR}$ & $N_{\rm SR}$ & $N_{\rm FR}$ &
$\Delta N_{\rm SR}$ & $\Delta N_{\rm FR}$ & $3\sqrt{N_{\rm NR}}$ & $d_3$ [kpc] \\
\midrule
Hyper-K  & IBD $\bar\nu_e$  & 10\,905 & 11\,042 & 9\,639 & $-137$ & $+1\,266$ & 312 & 40 \\
DUNE     & CC $\nu_e$+Ar    & 8\,428  & 8\,693  & 6\,870 & $-265$ & $+1\,558$ & 275 & 57 \\
JUNO     & IBD $\bar\nu_e$  & 1\,270  & 1\,286  & 1\,123 & $-16$  & $+147$    & 107 & 14 \\
\bottomrule
\end{tabular}
\end{table*}

\end{document}